\documentclass[a4paper]{article}

\usepackage{INTERSPEECH2019,amsmath,graphicx,multirow,booktabs,amssymb}

\title{Discriminative Learning for Monaural Speech Separation Using Deep Embedding Features}
\name{Cunhang Fan$^{1,3}$, Bin Liu$^1$, Jianhua Tao$^{1,2,3}$, Jiangyan Yi$^{1}$, Zhengqi Wen$^1$}
\address{
	$^1$NLPR, Institute of Automation, Chinese Academy of Sciences, Beijing, China\\
	$^2$CAS Center for Excellence in Brain Science and Intelligence Technology, Beijing, China\\
	$^3$School of Artificial Intelligence, University of Chinese Academy of Sciences, Beijing, China}
\email{\{cunhang.fan, liubin, jhtao, jiangyan.yi, zqwen\}\@nlpr.ia.ac.cn}

\begin{document}

\maketitle
\begin{abstract}
  Deep clustering (DC) and utterance-level permutation invariant training (uPIT) have been demonstrated promising for speaker-independent speech separation. DC is usually formulated as two-step processes: embedding learning and embedding clustering, which results in complex separation pipelines and a huge obstacle in directly optimizing the actual separation objectives. As for uPIT, it only minimizes the chosen permutation with the lowest mean square error, doesn't discriminate it with other permutations. In this paper, we propose a discriminative learning method for speaker-independent speech separation using deep embedding features. Firstly, a DC network is trained to extract deep embedding features, which contain each source's information and have an advantage in discriminating each target speakers. Then these features are used as the input for uPIT to directly separate the different sources. Finally, uPIT and DC are jointly trained, which directly optimizes the actual separation objectives. Moreover, in order to maximize the distance of each permutation, the discriminative learning is applied to fine tuning the whole model. Our experiments are conducted on WSJ0-2mix dataset. Experimental results show that the proposed models achieve better performances than DC and uPIT for speaker-independent speech separation. 
\end{abstract}

\section{Introduction}

Monaural speech separation aims to estimate target sources from mixed signals in a single-channel. It is a very challenging task, which is known as the cocktail party problem \cite{O2015Attentional}. 

In order to solve the cocktail party problem, many works have been done over the decades. Traditional speech separation methods include computational auditory scene analysis (CASA) \cite{wang2006computational}, non-negative matrix factorization (NMF) \cite{Schmidt2006Single} and minimum mean square error (MMSE) \cite{ephraim1985speech}. However, these methods have led to very limited success in speaker-independent speech separation \cite{cooke2010monaural}.

Recently, supervised methods using deep neural networks have significantly improved the performance of speech separation \cite{erdogan2018investigations,wang2018deep,luo2018tasnet,xu2018single,fan2018Utterance,wang2019apitch}. Deep clustering (DC) \cite{Hershey2016Deep} is a deep learning based method for speech separation and achieves impressive results. It trains a bidirectional long-short term memory (BLSTM) network to map the mixed spectrogram into an embedding space. At testing stage, the embedding vector of each time-frequency (TF) bin is clustered by K-means to obtain binary masks. However, the objective function of DC is defined in the embedding space, which can't be trained end-to-end. To overcome this limitation, the deep attractor network (DANet) \cite{chen2017deep} method is proposed. DANet creates attractor points in a high-dimensional embedding space of the acoustic signals. Then the similarities between the embedded points and each attractor are used to directly estimate a soft separation mask at the training stage. Unfortunately, it enables end-to-end training while still requiring K-means at the testing stage. In other words, it applies hard masks at testing stage.

The permutation invariant training (PIT) \cite{Yu2017Permutation} and utterance-level PIT (uPIT) \cite{Kolbaek2017Multitalker} are proposed to solve the label ambiguity or permutation problem of speech separation. The PIT method solves this problem by minimizing the permutation with the lowest mean square error (MSE) at frame level. However, it does not solve the speaker tracing problem. To solve this problem, uPIT is proposed. With uPIT, the permutation corresponding to the minimum utterance-level separation error is used for all frames in the utterance. Therefore, uPIT doesn't need speaker tracing step during inference. However, uPIT and PIT only use the mixed amplitude spectrum as input features, which can't discriminate each speaker very well. In addition, uPIT doesn't increase the distance between the chosen permutation and others. This may lead to increasing the possibility of remixing the separated sources. In \cite{wang2018alternative}, a Chimera network \cite{luo2017deep} is applied to speech separation, which uses a multi-task learning architecture to combine the DC and uPIT.  However, it simply employs the DC and uPIT as two outputs rather than fusion with each other.

In this paper, in order to address the problems of DC and uPIT, we propose a discriminative learning method for speaker-independent speech separation with deep embedding features. uPIT is incorporated into DC-based speech separation framework. Firstly, a DC network is trained to extract deep embedding features. Clusters in the embedding space can represent the inferred spectral masking patterns of individual source. Therefore, these deep embedding features contain the information of each source, which is conducive to speech separation. Then instead of using K-means clustering to estimate hard masks, we make full use of the uPIT network to directly learn each source's soft mask from the embedding features. And then uPIT and DC are jointly trained, which directly optimizes the actual separation objectives. Finally, in order to decrease the possibility of remixing the separated sources,  motivated by our previous work \cite{fan2018Utterance}, we apply the discriminative learning to fine tuning the separated model. 

The rest of this paper is organized as follows. Section 2 presents the single channel speech separation based on masks. The  proposed method is stated in section 3. Section 4 shows detailed experiments and results. Section 5 draws conclusions.

\section{Single Channel Speech Separation}
\label{sec:format}

The object of single channel speech separation is to separate target sources from a mixed signal.
\begin{equation}
y(t) = \sum_{s=1}^{S}{x_s(t)}
\label{eq1}
\end{equation}
where \(y(t)\) is the mixed speech, \(S\) is the number of source signals and \(x_s(t)\), \(s=1,...,S\) are target sources. The corresponding short-time Fourier transformation (STFT) of those signals are \(Y(t,f)\) and \(X_s(t,f)\). 


Our aim is to estimate each source signal \(x_s(t)\) from \(y(t)\)  or \(Y(t,f)\) . It is well-known that mask based speech separation can obtain a better result \cite{Wang2014On}. According to the commonly used masking method, the estimated magnitude  \(|\widetilde{X}_s(t,f)|\)  of each source can be estimated by
\begin{equation}
|\widetilde{X}_s(t,f)|=|Y(t,f) |\odot{M_s(t,f)}
\label{eq3}
\end{equation}
where \(\odot\) indicates element-wise multiplication and \(M_s(t,f)\) is the mask of source \(s\). It is very difficult to estimate phase directly for speech separation and speech enhancement. Therefore, the estimated magnitude  \(|\widetilde{X}_s(t,f)|\) and the phase of mixed signals are utilized to reconstruct each source signal by inverse STFT (ISTFT).

\section{The Proposed Speech Separation System}
\label{sec:pagestyle}

In this section, we present our proposed discriminative learning method for speaker-independent speech separation with deep embedding features, which is shown in Figure~\ref{fig:DC_uPIT_system}. From DC network \cite{Hershey2016Deep} we can know that clusters in the deep embedding space can represent the inferred spectral masking patterns of individual sources. Therefore, these embedding vectors are discriminative features for speech separation. Motivated by this, we use this deep embedding vectors as the input of separation system. Then a uPIT network is used to learn the soft mask of each source instead of K-means clustering. Moreover, in oder to maximize the distance of each speaker, the discriminative learning is applied to fine tuning the whole model. Finally, uPIT and DC are jointly optimized.

\subsection{Deep embedding features}

\begin{figure}[t]
	\centering
	\includegraphics[width=\linewidth]{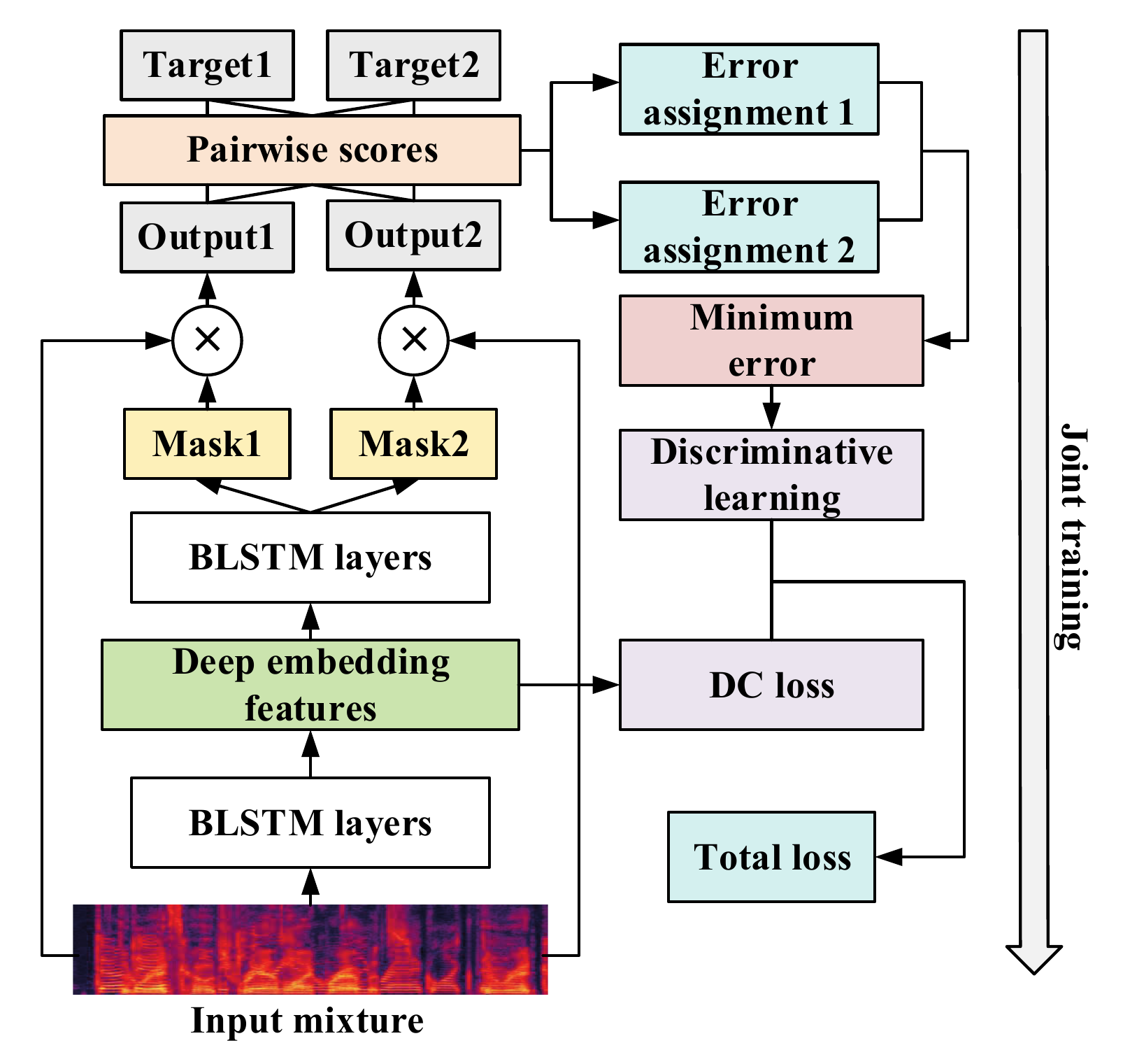}
	\caption{Schematic diagram of our proposed DL-DEF speech separation system. DC loss is the loss of deep clustering.}
	\label{fig:DC_uPIT_system}
\end{figure}

As shown in Figure~\ref{fig:DC_uPIT_system}, we firstly train a BLSTM network as the extractor to acquire deep embedding features. The input of BLSTM is the mixed amplitude spectrum \(|Y(t,f)|\) and the output is the D-dimensional deep embedding features V.
\begin{equation}
V=f_{\theta}(|Y(t,f)|) \in{\mathbb{R}^{TF\times{D}}}
\label{eq4}
\end{equation}
where TF is the number of T-F bins and \(f_{\theta}(*)\) is a mapping function based on the BLSTM network.

The loss function of embedding features' network is defined as follow:
\begin{equation}
\begin{split}
J_{DC} &=||VV^T-BB^T||_F^2\\
&=||VV^T||_F^2-2||V^TB||_F^2+||BB^T||_F^2\\
\end{split}
\label{eq5}
\end{equation}
where \(B\in{\mathbb{R}^{TF\times{C}}}\) is the source membership function for each T-F bin, i.e.,\(B_{tf,c}=1\), if source \(c\) has the highest energy at time \(t\) and frequency \(f\) compared to the other sources. Otherwise, \(B_{tf,c}=0\). C is the number of source. \(||*||_F^2\) is the squared Frobenius norm. \(J_{DC}\) is the DC loss in Figure~\ref{fig:DC_uPIT_system}.

\subsection{Speech separation model based on deep embedding features}

Different from deep clustering \cite{Hershey2016Deep} utilizing the K-means clustering to acquire hard masks, we use the embedding vectors as the input of uPIT to directly learn each source's soft masks. Therefore, the DC and uPIT can be trained end-to-end. When these features are extracted, they are reshaped as \(V'\in{\mathbb{R}^{T\times{FD}}}\). Then these embedding features \(V'\) are sent to the separated system. In this way, we can learn a soft mask of each source from these features, which is better than the hard mask estimated by K-means clustering.

In order to make full use of phase information, we use the ideal phase sensitive mask (IPSM) \cite{Erdogan2015Phase} in this work. The IPSM is defined as
\begin{equation}
M_s(t,f)=\frac{|X_s(t,f)|cos(\theta_y(t,f)-\theta_s(t,f))}{|Y(t,f)|}
\label{eq7}
\end{equation}
where \(\theta_y(t,f)\) and \(\theta_s(t,f)\) are the phase of mixed speech and target source.

During the separation stage, the uPIT is used to estimate each source. We apply the MSE between estimated magnitude and true magnitude as the training criterion.
\begin{equation}
J_{\phi_p(s)}=\frac{1}{TF}\sum_{s=1}^S|||Y|\odot{\widetilde{M}_s}-|X_{\phi_p(s)}|cos(\theta_y-\theta_s)||_F^2
\label{eq8}
\end{equation}
where \(\widetilde{M}_s\) is the estimated mask. \(\phi_p(s)\), \(p\in{[1,P]}\) is an assignment of target source s, \(P=S!\) (\(!\) denotes the factorial symbol) is the number of permutations.

In order to solve the label ambiguity problem, the minimum cost among all permutations (P) is chosen as the optimal assignment.
\begin{equation}
\phi^*=\mathop{arg\,min}_{\phi\in{P}}J_{\phi_p(s)}
\label{eq9}
\end{equation}

\subsection{Discriminative learning}

For uPIT, the target of minimizing Eq.\ref{eq9} is to make the predictions and targets more similar. In this paper, we explore discriminative objective function that not only increase the similarity between the prediction and its target, but also decrease the similarity between the prediction and the interference sources.

The discriminative learning maximizes the dissimilarity between the chosen permutation \(\phi^*\) and the other permutations by adding a regularization at the cost function. The cost function of the proposed model is defined as
\begin{equation}
J_{DL}=\phi^*-\sum_{\phi \ne{\phi^*},\phi \in{P}}\alpha \phi
\label{eq10}
\end{equation}
where \(\phi\) is a permutation from \(P\) but not \(\phi^*\), \(\alpha \ge0\) is the regularization parameter of \(\phi\). When \(\alpha=0\), there is no discriminative learning, which is same as the loss of uPIT.

For two-talker speech separation, we assume that \(\phi_1\) is the permutation with the lowest MSE. Therefore, the cost function becomes as follow:
%
%
\begin{equation}
\begin{split}
J_{DL} &=\phi_1-\alpha{\phi_2}\\
&=\frac{1}{TF}\sum(|||Y|\odot{\widetilde{M}_1}-|X_1|||_F^2-\alpha{|||Y|\odot{\widetilde{M}_1}-|X_2|||_F^2}\\
&+|||Y|\odot{\widetilde{M}_2}-|X_2|||_F^2-\alpha{|||Y|\odot{\widetilde{M}_2}-|X_1|||_F^2})\\
\end{split}
\label{eq11}
\end{equation}

From Eq.\ref{eq11} we can know that the discriminative learning enlarges the distance of the target source with the interference sources. It means that it maximizes the differences between the target speakers with the others. 

Therefore, the proposed model with discriminative learning minimizes the distance between the outputs of model and their corresponding reference signals. Simultaneously, it maximizes the dissimilarity between the target source and the interference. So the discriminative learning decreases the possibility of remixing the separated sources.

\subsection{Joint training loss function}
\label{ssec:Multi-task}

The deep clustering objective can reduce within-source variance in the internal representation \cite{luo2017deep}. Therefore, in order to effectively extract embedding features, we make full use of a joint training framework for the proposed system. More specifically, the deep clustering objective is added at the loss function.
\begin{equation}
\begin{split}
J & =\lambda{J_{DC}}+(1-\lambda)J_{DL}\\
& = \lambda{J_{DC}}+(1-\lambda)(\phi^*-\sum_{\phi \ne{\phi^*},\phi \in{P}}\alpha \phi)
\end{split}
\label{eq12}
\end{equation}
where \(J\) is the joint training loss function of the proposed system. \(\lambda\in{[0,1]}\) controls the weight of two objectives. Note that when \(\lambda=1\), the proposed method is same as deep clustering \cite{Hershey2016Deep}.





In order to get the better deep embedding features, we only train the DC network firstly. The loss function is Eq.\ref{eq5}. Then these deep embedding features are used to train the separated model based on uPIT. The loss function is with no discriminative learning:
\begin{equation}
J^{\prime}=\lambda{J_{DC}}+(1-\lambda){\phi^*}
\label{eq13}
\end{equation}
Finally, we apply the discriminative learning to fine tuning the whole model by the joint training loss function \(J\) in Eq.\ref{eq12}.



\section{Experiments and Results}
\label{sec:typestyle}

\subsection{Dataset}
\label{ssec:subhead1}

Our experiments are conducted on the WSJ0-2mix dataset \cite{Hershey2016Deep}, which is derived from WSJ corpus \cite{garofalo2007csr}. The WSJ0-2mix dataset consists three sets: training set (20,000 utterances about 30 hours), validation set (5,000 utterances about 10 hours) and test set (3,000 utterances about 5 hours). Specifically, training and validation set are generated by randomly selecting utterances from WSJ0 training set (\texttt{si\_tr\_s}) with signal-to-noise ratios (SNRs) between 0dB and 5dB. Similar as generating training and validation set, the test set is created by mixing the utterances from the WSJ0 development set (\texttt{si\_dt\_05}) and evaluation set (\texttt{si\_et\_05}). 

We use the validation set to evaluate the source separation performance in closed conditions (CC). Moreover, because the speakers in the test set are different from those in the training set and validation set, the test set is considered as open condition (OC) evaluation.

\subsection{Experimental setup}

The sampling rate of all generated data is 8 kHz before processing to reduce computational and memory costs. The 129-dim normalized spectral magnitudes of the mixed speech are used as the input features, which are computed using a short-time Fourier transform (STFT) with 32 ms length hamming window and 16 ms window shift. The magnitudes of two targets are generated in the same way. Our models are implemented using Tensorflow deep learning framework \cite{Abadi2016TensorFlow}.

In this work, the deep embedding network has two BLSTM layers with 896 units. The embedding dimension D is set to 40. A tanh activation function is followed by the embedding layer. As for the separated network, it has only one BLSTM layer with 896 units. Therefore, there are three BLSTM layers in this work, which keeps the network configuration the same as baseline in \cite{Kolbaek2017Multitalker}. A Rectified Liner Uint (ReLU) activation function is followed by the uPIT network, which is the mask estimation layer. The regularization parameter \(\alpha\) of discriminative learning is set to 0.1.

\setlength{\tabcolsep}{1.5mm}{
\begin{table*}[t]
	\caption{The results of SDR, SIR, SAR and PESQ for different separation methods on closed (CC) and open (OC) condition. \(\lambda\) is the weight of joint training in Eq.\ref{eq12} and \ref{eq13}. DEF denotes the deep embedding features. uPIT is the baseline method, uPIT+DEF and uPIT+DEF+DL are our proposed methods. uPIT+DEF means with no discriminative learning.}
	\label{tab:results1}
	\centering
	\begin{tabular}{c |c|cc|cc|cc|cc|cc|cc|cc|cc}
		\toprule
		\multicolumn{1}{c}{\multirow{3}*{Method}} & \multicolumn{1}{|c|}{\multirow{3}*{\(\lambda\)}} & \multicolumn{8}{|c|}{Optimal (Opt.) Assign.} & \multicolumn{8}{|c}{Default (Def.) Assign.} \\
		\cline{3-18}
		& & \multicolumn{2}{|c|}{SDR(dB)} & \multicolumn{2}{|c|}{SIR(dB)} & \multicolumn{2}{c}{SAR(dB)} & \multicolumn{2}{|c|}{PESQ} & \multicolumn{2}{|c|}{SDR(dB)} & \multicolumn{2}{|c|}{SIR(dB)} & \multicolumn{2}{c}{SAR(dB)} & \multicolumn{2}{|c}{PESQ}\\
		\cline{3-18}
		& & CC& OC& CC& OC& CC& OC & CC& OC& CC& OC& CC& OC& CC& OC& CC& OC\\
		\midrule
		uPIT(baseline)&     -&   11.3& 11.2&  18.8& 18.8& 12.3& 12.3&  2.68&   2.67&  10.3 & 10.1& 17.7 & 17.5& 11.5 & 11.3& 2.60& 2.58\\
		\midrule
		uPIT+DEF&  0.01& 11.7& 11.6& 19.4& 19.5& 12.7& 12.6& \textbf{2.85}& \textbf{2.84}& 10.8 & 10.7&  18.4& 18.4 & \textbf{12.0}&  11.8& \textbf{2.77}&{2.75}\\
		uPIT+DEF&  0.05& 11.7& 11.7& 19.5& 19.6& 12.7& 12.6& {2.84}& \textbf{2.84}&   10.8& \textbf{10.8}& 18.4& \textbf{18.8}& 11.9& \textbf{11.9} & {2.76}&{2.75}\\
		uPIT+DEF&  0.1& 11.7& 11.7& 19.5& 19.5& 12.7& 12.6& {2.84}& \textbf{2.84}&   10.8& 10.7& 18.5& 18.4& \textbf{12.0}& \textbf{11.9} & 2.76&2.74\\
		\midrule
		uPIT+DEF+DL&  0.05& \textbf{11.9}& \textbf{11.9}& \textbf{19.9}& \textbf{20.0}& \textbf{12.8}& \textbf{12.7}& 2.83& 2.83&   \textbf{11.0}& \textbf{10.8}& \textbf{18.8}& \textbf{18.8}& \textbf{12.0}& \textbf{11.9} & 2.74 &2.73\\
		\bottomrule
	\end{tabular}
\end{table*}}

All models contain random dropouts with a dropout rate 0.5. Each minibatch contains 16 randomly selected utterances. The minimum number of epoch is 30. The learning rate is initialized as 0.0005 and scaled down by 0.7 when the training objective function value increased on the development set. The early stopping criterion is that the relative loss improvement is lower than 0.01. Our models are optimized with the Adam algorithm \cite{Kingma2014Adam}.

\subsection{Baseline model and evaluation metrics}

We re-implement uPIT with our experimental setup as our baseline. It has three BLSTM layers with 896 units. The others are same as our experimental setup. In this work, in order to quantitatively evaluate speech separation results, the models are evaluated on the signal-to-distortion ratio (SDR), signal-to-interference ratio (SIR) and signal-to-artifact ratio (SAR) which are the BBS-eval \cite{vincent2006performance} score, and the perceptual evaluation of speech quality (PESQ) \cite{rix2002perceptual} measure. 

\subsection{Experimental results}

Table~\ref{tab:results1} shows the results of SDR, SIR, SAR and PESQ between the proposed method and uPIT-BLSTM on the WSJ0-2mix database. DEF denotes the deep embedding features.



\subsubsection{Evaluation of deep embedding features}

From Table~\ref{tab:results1}, we can find that our proposed uPIT+DEF methods outperform baseline uPIT in all objective measures no matter optimal assignment (Opt. assign.) or default assignment (Def. assign.). These indicate that the deep embedding features are more easily separated than the mixed amplitude spectral features. Because deep embedding features contain the potential masks of individual sources and they can effectively discriminate different target speakers. Therefore, they are conducive to speech separation.

Moreover, in order to acquire better deep embedding features, we propose a novel joint training framework to instruct the training of deep embedding network. Three different weights \(\lambda\) (0.01, 0.05 and 0.1) are applied. From Table~\ref{tab:results1}, we can know that the performance of these three \(\lambda\) for speech separation are similar. The reason is that we firstly train a DC network with 30 epochs, which can obtain a pretty good representation for deep embedding features. Therefore, these three \(\lambda\) get similar performance. 

\subsubsection{Evaluation of discriminative learning}

Since the discriminative learning separates the target speaker with others, it provides a constraint to ensure that the output frames of the same speaker do not remix to the interferences. Therefore, we use discriminative learning to fine tuning the whole model based on \(\lambda=0.05\). From Table~\ref{tab:results1}, we can know that when the discriminative learning is applied, our proposed method uPIT+DEF+DL achieves better performances than the proposed uPIT+DEF overall objective scores, except for the PESQ measure. This shows the effectiveness of the discriminative learning. Meanwhile, compared with the uPIT baseline system, the proposed uPIT+DEF+DL gets a better performance in all cases. For example, as for the optimal assignment on open condition, the proposed method achieves 6.3\%, 6.4\% and 3.3\% relative improvements in SDR, SIR and SAR over the uPIT baseline system. These results reveal the effectiveness of our proposed method.

\subsubsection{Comparisons with other separation methods}

In order to better compare the performance of our proposed method (uPIT+DEF+DL) and other separation methods, Table~\ref{tab:results2} presents the results of SDR (dB) in the other competitive approaches on the same WSJ0-2mix dataset. Note that, for \cite{xu2018single,Hershey2016Deep,xu2018shifted,Kolbaek2017Multitalker,isik2016single,chen2017deep} methods are use SDR improvements results. Therefore, we manually add 0.2 dB to their final results although the SDR result of the mixture is only about 0.15 dB. Compared with other speech separation methods, our proposed method improves the performance significantly to 11.9 dB and 10.8 dB with no phase enhancement for Opt Assign and Def Assign. Moreover, from Table~\ref{tab:results2}, we can know that our proposed method outperforms other methods, such as DC+, DANet, SDC-MLT-Grid LSTM. Compared with DC+ \cite{isik2016single}, our proposed method achieves 12.5\% relative improvement on open condition. These results confirm that using discriminative learning and deep embedding features can improve the performance of speaker-independent speech separation.

\begin{table}[t]
	\caption{The results of SDR (dB) in the other different separation methods on the WSJ0-2mix dataset on CC and OC with no phase enhancement.}
	\label{tab:results2}
	\centering
	\begin{tabular}{|c|c|c|c|c|}
		\hline
		\multirow{2}{*}{Method}   & \multicolumn{2}{c|}{Opt Assign} & \multicolumn{2}{c|}{Def Assign} \\ \cline{2-5} 
		& CC             & OC             & CC             & OC             \\ \hline
		DC\cite{Hershey2016Deep}                        & -              & -              & 6.1            & 6.0            \\ 
		DC+\cite{isik2016single}                       & -              & -              & -              & 9.6            \\ 
		DANet\cite{chen2017deep}                     & -              & -              & -              & 9.8            \\ 
		uPIT-BLSTM \cite{Kolbaek2017Multitalker}               & 11.1           & 11.0           & 9.6            & 9.6            \\
		cuPIT-Grid LSTM-RD\cite{xu2018single}        & 11.4           & 11.4           & 10.4           & 10.3           \\
		SDC-MLT-Grid LSTM \cite{xu2018shifted}        & 11.6           & 11.6           & 10.8           & 10.7           \\ \hline
		uPIT+DEF+DL(our proposed) & \textbf{11.9}           & \textbf{11.9}          & \textbf{11.0}           & \textbf{10.8}           \\ \hline
	\end{tabular}
\end{table}


\section{Conclusions}

In this paper, we propose a speaker-independent speech separation method with discriminative learning based on deep embedding features. We firstly train an DC network to extract deep embedding features. Then these features are used as the input of uPIT system to directly separate the different speaker sources. Moreover, uPIT and DC  are jointly optimized. Finally, the discriminative learning is applied to fine tuning the whole model. Results show that the proposed method outperforms uPIT baseline, with a relative improvement of 6.3\%, 6.4\% and 3.3\% relative improvements in SDR, SIR and SAR, respectively. In the future, we will explore phase enhancement based on the proposed method.

\section{Acknowledgements}

This work is supported by the National Key Research \& Development Plan of China (No.2017YFB1002802), the NSFC (No.61425017, No.61831022, No.61771472, No.61603390), the Strategic Priority Research Program of Chinese Academy of Sciences (No.XDC02050100), and Inria-CAS Joint Research Project (No.173211KYSB20170061). Authors also thank Shuai Nie for his helpful comments on this work.

\vfill\pagebreak

\bibliographystyle{IEEEtran}

\bibliography{mybib}


\end{document}